\documentclass[dvips,includegraphics]{article}
\usepackage{icrctc07}

\title{Search for very high energy gamma rays from possible ultra-high 
energy cosmic ray sources by the MAGIC Telescope}

\shorttitle{Search for VHE gamma rays from possible UHE cosmic ray 
sources by the MAGIC Telescope}

\authors{Kenji Shinozaki and Masahiro Teshima on behalf of the 
MAGIC Collaboration}

\shortauthors{K. Shinozaki and M. Teshima}

\afiliations{Max-Planck-Institut f\"ur Physik, F\"ohringer Ring 6, 
D-80805 M\"unchen, Germany}
\email{kenjikry@mppmu.mpg.de}

\abstract{
The origin of ultra-high energy (UHE) cosmic rays is still an open 
question. In the present work, we searched the possible UHE cosmic 
ray sources using the MAGIC telescope for the associated very high 
energy (VHE) gamma ray emission. Due to constrained propagation 
distance of such cosmic rays, we selected nearby galaxies in vicinity 
of the direction of the AGASA triplet and a HiRes UHE cosmic ray event: 
NGC3610 and NGC3613 (quasar 
remnants); Arp299 (a system of colliding galaxies). No significant 
excess in the VHE region was found found from these objects or 
their surrounding region. At multi-100 GeV regime, the upper limits 
on fluxes were given against gamma ray sources in surrounding region. 
The presented limits constrain the flux of a new hypothetical source
in the region, provided the cosmic rays are emitted from a single
point-like origin.}

\begin{document}
\maketitle

\section{Introduction}

Cosmic rays up to an $\sim 3\times 10^{20}$ eV energy have been 
observed so far, while their origin remains unidentified. In the standard 
particle acceleration models, possible candidates are only a few types of 
the most powerful or very large-scale objects. Due to the 
lack of data on Galactic and extragalactic magnetic fields, it is 
difficult to trace back to their sources. However, the AGASA group 
claimed that a part of UHE cosmic ray events cluster in their arrival 
direction distribution (\cite{ags4} and references therein). Several 
groups have also pointed out a notable angular correlation with 
particular class of objects such as BL Lacertae 
objects, which include known VHE gamma ray emitters \cite{bll}. A similar 
correlation was reported on quasar remnants \cite{dq}. By the quasar 
remnant scenario \cite{tev_uhe}, the rotational energy of the black 
hole is dominantly radiated in MeV--TeV band whose luminosity can be a 
few orders higher than that of cosmic rays. If quasar remnants within 
several tens of Mpc are the origin of all observed cosmic rays above 
$10^{20}$ eV, they would overshine more than any known TeV gamma ray 
source. But in the case that hundreds of such sources distribute over 
several 100 Mpc space, the cosmic ray flux, at least around several 
$10^{19}$ eV, is consistent with flux limits by VHE gamma ray 
observations.  

\section{Observation}

To find an evidence or hint of the UHE cosmic ray
source, especially of the clustering events, we searched 
possible sources for VHE gamma rays with MAGIC (Major Atmospheric 
Gamma ray Imaging Cherenkov) Telescope  \cite{magic}. The detector 
consists of the world's largest 17-m-diameter reflector viewed by 
photo-multiplier tube camera with 3.5$^\circ$ field of view 
(FOV). It is located at the Observatorio del Roque de los Muchachos
(2200 m above sea level), La Palma, Canary Islands. 

Table~\ref{tab1} summarises the UHE cosmic ray cluster (AGSAA triplet) 
in the Plough, Ursa Major where with AGASA data alone three events 
above $4\times 10^{19}$ eV had been observed in a 2.5$^{\circ}$ radius.  
Recently, the HiRes group detected a $3.7\times 10^{19}$ eV near the triplet 
and estimated the chance probability of observing such `quartet' to be 
0.6\% \cite{hr}. This direction coincides in the supergalactic plane. 

Along the direction of interest, we selected nearby objects that are 
capable sources of UHE cosmic ray observed on the Earth. To carry 
out delegated observations, we picked out the following three objects 
as summarised in Table~\ref{tab2}.

NGC3610 is a merger remnant elliptical galaxy with a black hole of 
$\sim 0.5 \times 10^{8}$ solar masses $M_\odot$. The maximum accessible 
cosmic ray energy is estimated to be $4.4 \times 10^{19}$ eV. This galaxy 
has attracted attentions because of its fine structure and prominently 
warped disk, implying that a dynamical event occurred in a few $10^{9}$ 
years ago \cite{dis3610}. 

NGC3613 is a quasar remnant with a black hole of 
$1.6 \times 10^{8} M_\odot$. The maximum cosmic ray energy is 
estimated to be $5.9 \times 10^{19}$ eV. This galaxy is 
located 0.7$^\circ$ off the most energetic triplet event.

Arp299 (Mrk 171, VV118; \cite{x299} and references therein) is 
a starburst galaxy and a system of colliding galaxies 
(NGC3690+IC694). It is characterised by an 
extreme star formation rate. The supernova rate is also
as high as $\sim 0.5$ per year. Arp299 is pointed out to 
be a potential source to explain the AGASA triplet \cite{gil}. 

Operating the MAGIC telescope, we observed NGC3610 for 5.1 hours, 
NGC3613 for 5.0 hours each in December 2006 and Arp299 for 9.9 
hours in January 2007. All the observations were carried out by 
the so-called wobble mode \cite{wobble} in which the object position 
was tracked alternatively 0.4$^\circ$  east or west off the centre of FOV. 

\begin{table}[t]
\begin{center}
\begin{tabular}{ccccc}
\hline
Dataset & $E_0^{{\rm (CR)}}$ & R.A. & Decl. & $\delta\theta$ \\
\hline
AGASA   & 77.6 & 11h14m & $+57.6^\circ$ & $1.0^\circ$  \\
AGASA   & 55.0 & 11h29m & $+57.1^\circ$ & $1.7^\circ$  \\
AGASA   & 53.5 & 11h13m & $+56.0^\circ$ & $1.9^\circ$  \\
HiRes   & 37.6 & 11h16m & $+55.85^\circ$ & $0.4^\circ$  \\
\hline
\end{tabular}
\caption{UHE events in the cluster. $E_0$ is the estimated 
energy by each experiment in $10^{18}$ eV (EeV). 
$\delta\theta$ is a nominal direction error at a 68\%CL.} \label{tab1}
\end{center}
\end{table}

\begin{table}[t]
\begin{center}
\begin{tabular}{ccccc}
\hline
Object & R.A. & Decl. & $m_{\rm V}$ & $D$  \\
\hline
NGC3610 & 11h18m4 & $+58^\circ$47'0 &  8.5 & 31 \\
NGC3613 & 11h18m6 & $+58^\circ$00'6 &  8.9 & 34 \\
Arp299  & 11h28m5 & $+58^\circ$33'9 & 11.8 & 37 \\
\hline
\end{tabular}
\caption{Locations of the observed objects in the present work.
$m_{V}$ is visible magnitude. $D$ is distance in Mpc ($H_0 \sim 70$ 
[km h$^{-1}$ Mpc$^{-1}$]).}\label{tab2}
\end{center}
\end{table}

\section{Analysis and result}

The data analysis was performed by our standard analysis 
chain MARS (MAGIC Analysis and Reconstruction Software) \cite{mars}. 
The image of observed showers was parameterised by the conventional
Hillas technique \cite{hillas}. To interpret data, a number of simulated 
air showers were generated under actual telescope configurations 
\cite{mmcs}. For rejection of the background hadronic shower events, 
we used the random forest (RF) method \cite{rf}. To define the parameter 
`hadronness' $H$ that represents how hadron-like showers look (=0 for 
gamma ray- and =1 for hadron- like), the RF algorithm is trained 
with data and simulated gamma ray shower samples for compatible 
zenith angles $(29^\circ-45^\circ)$. The energy of primary gamma rays
was estimated similarly by the RF with simulated 
gamma ray showers. The energy resolution is $< 25\%$ at 
energies of interest. The incoming direction of showers was 
reconstructed by the so-called DISP method \cite{disp3}. 
The typical angular resolution is $\sim 0.1^\circ$.

First we search for the VHE gamma ray emission directly from the 
observed object. To select gamma ray like showers, the criterion 
of $H$ cut was optimised by simulated gamma ray showers by maximizing 
their significance against surviving background events. $\theta^2$ 
distribution is compared with that of OFF-sources 
(expected background distribution) where 
$\theta$ is the angular distance between 
reconstructed shower incoming direction and the object position. 
The cut for $\theta^2$, typically $\sim 0.02$ [degree$^2$], was 
similarly optimised by simulated showers and was applied on the data.  

\begin{figure}[t]
\begin{center}
\noindent
\includegraphics [width=.43\textwidth]{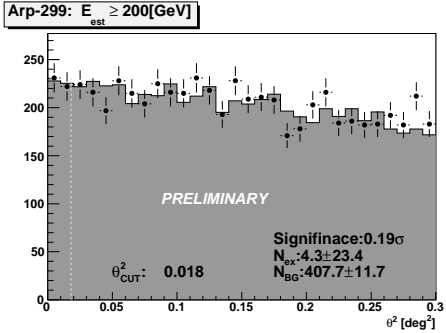}
\end{center}
\caption{An example of $\theta^{2}$ distribution (Arp299; 
$E_{\rm est}\ge 200$[GeV]) . Closed circles: ON-source
Histogram: OFF-source data, respectively. The vertical 
line: optimised cut of $\theta^2$ for this case}
\label{fig1}
\end{figure}

Figure~\ref{fig1} shows an example of $\theta^2$ distribution on Arp299
($E_{\rm est} \ge 200$ [GeV]) in which no significant excess was found
for gamma ray signals. 

Also for other objects, no significant excesses was found. 
Therefore the upper limits of gamma ray fluxes were estimated for 
these cases. In four estimated energy bins, the 
acceptance of gamma ray showers after cuts were evaluated by the simulated 
samples independent of ones used in the RF.

The limits on the fluxes at a 95\% confidence 
level (CL) are summarised in Table~3. The integral flux limits above 
$\ge 200$ GeV correspond to $\sim 7\%$ for NGC 3610 and NGC 3613 and 
$5\%$ for Arp 299 to the Crab flux observed by MAGIC \cite{crab}.

\begin{table}
\begin{center}
\begin{tabular}{cccc}
\hline
$\langle E_0 \rangle$ & \multicolumn{3}{c}{UL 
[$10^{-14}$ cm$^{-2}$ s$^{-1}$ GeV$^{-1}$]} \\
\cline{2-4}
 [GeV]  & NGC3610 & NGC3613 & Arp299 \\
\hline
216 &  18. & 16. & 12.\\
463 &  2.5  &  2.3 &  1.7 \\
921 &  0.61 &  0.60 & 0.43\\
1266 & 0.32 & 0.30 &0.24\\
\hline
\end{tabular}
\end{center}
\caption{Upper limits on gamma ray fluxes from observed objects at 
a 95\% CL.}
\end{table}

\begin{figure}[t]
\begin{center}
\noindent
\includegraphics [width=.43\textwidth]{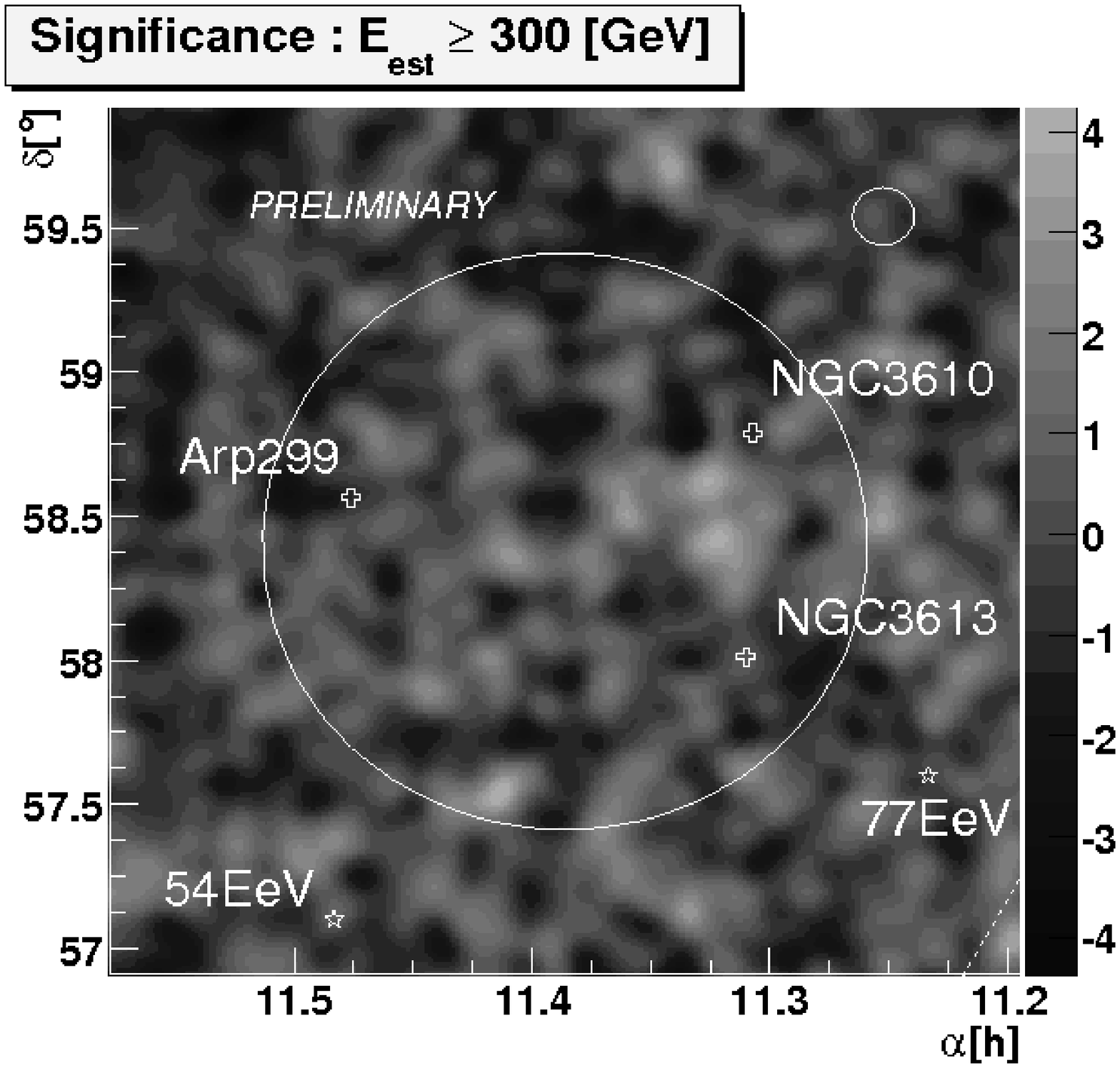}
\end{center}
\caption{Significance map $(E_{\rm est} \ge 300 [{\rm GeV}])$ of the 
observed region smoothed by the angular resolution. 
$\sim 0.1^\circ$. 
Crosses: positions of objects. Stars: AGASA 
events. The solid curve: $1^\circ$ circle from the 
cicumcentre of observed objects.}
\label{fig2}

\begin{center}
\noindent
\includegraphics [width=.43\textwidth]{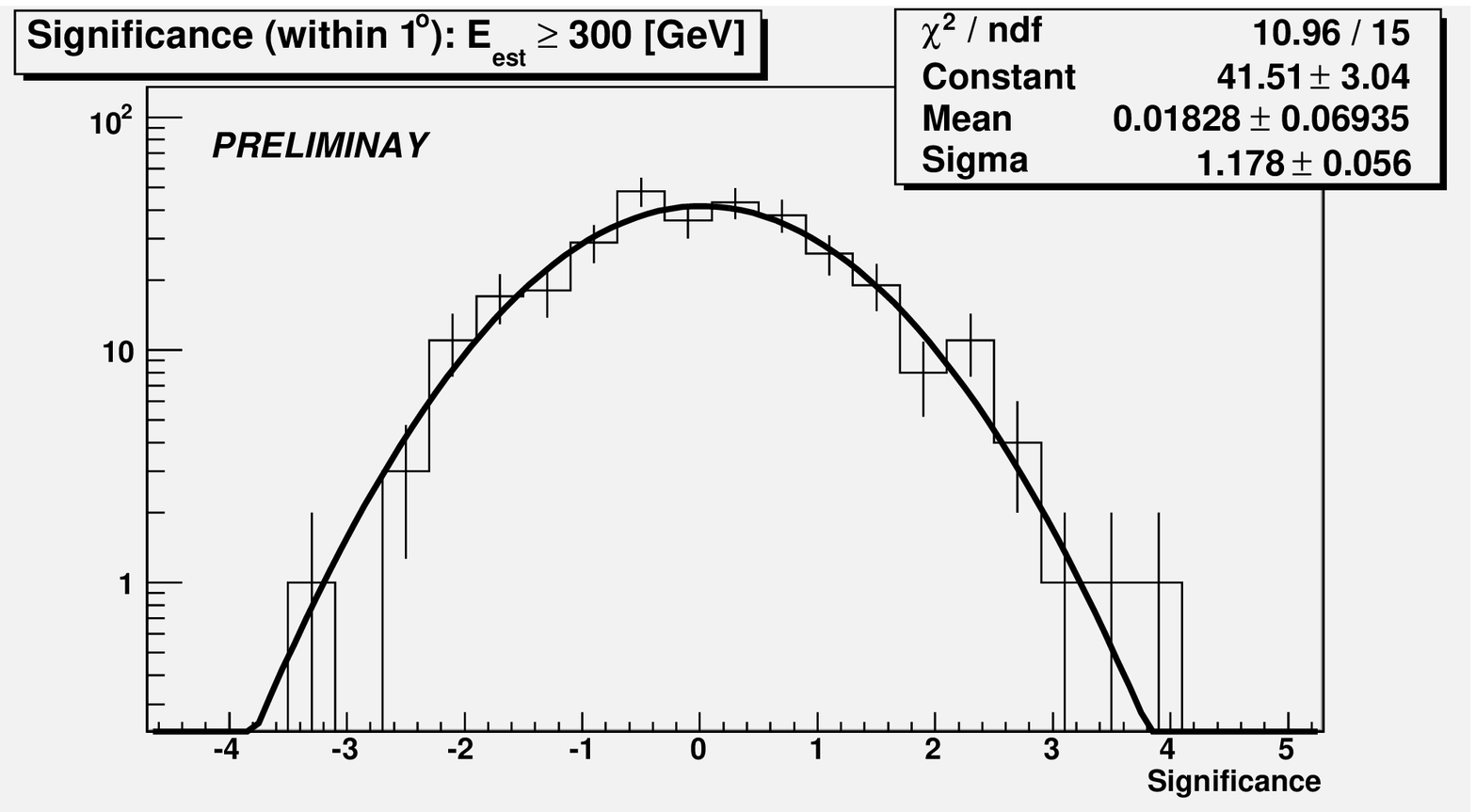}

\end{center}
\caption{The significance distribution within 
solid curves ($1^\circ$ radius) in the map of 
Figure~\ref{fig2}.}
\label{fig3}
\end{figure}

To search for any emission apart from the thse objects, the 
significance of the excess  
events by Equation (17) in \cite{lima} was estimated 
for each sky point using the cumulative dataset. The 
background distribution was modelled by the data
observed on the non-source region at similar zenith
angles.

Figure~\ref{fig2} shows the significance map for 
$E_{\rm est} \ge 300$ GeV with a convolution of the angular resolution. 
Crosses denote positions of objects. Stars represent two  
AGASA events. The solid curve is a $1^\circ$ circle centred 
on the cicumcentre of three objects to accomodate to a moderately
flat sensitivity.
Figure~\ref{fig3} shows the significance distribution for
$0.1^\circ\times 0.1^\circ$ bins within this circle. The distribution is compatible with non-source 
hypothesis.

\section{Concluding remarks}

Following up with AGASA-HiRes quartet detection, we searched the 
possible UHE cosmic ray sources, nearby quasar remnants NGC3610 
and NGC3613 and a starburst galaxy Arp299, for VHE emission. 
In $\sim 200 - 500$ GeV 
energies, the upper limits on the gamma ray flux  from each source 
is placed against each object and were $\sim 8\%-12\%$ Crab flux at 
a 95\% CL.

Over the region observed, there are no positive signals for the VHE 
emission by effectively $\sim 15$ hour observation.  Assuming the 
triplet is a
UHE cosmic ray signal from a single source, its energy flux yields 
$\sim 2$ eV cm$^{-2}$ s$^{-1}$ by the 
AGASA observation. If any of the observed objects is a responsible 
cosmic ray source, the present limits correspond to $\sim 3-5$ times of 
the energy flux of UHE cosmic ray component. 

In near future, if nearby sources exist, UHE cosmic ray clusters will 
be found even clearly by higher quality data provided by 
$> 1000$-km$^2$-scale observatories. With progress of imaging Cherenkov
telescopes as well, it is highly expected to identify the sources by both 
ways of cosmic rays physics and gamma ray astronomy to approach the the 
mystery of UHE cosmic ray origin.

\section*{Acknowledgment}

The authors appreciate the excellent working conditions at the 
ORM. They acknowledge the supports by German BMBF and MPG, Italian 
INFN and Spanish CICYT, ETH research grant TH 34/04 3, and the Polish 
MNiI grant 1P03D01028.  KS was supported by Japanese JSPS.

\end{document}